\documentclass[manuscript,screen]{acmart}
\AtBeginDocument{%
  \providecommand\BibTeX{{%
    \normalfont B\kern-0.5em{\scshape i\kern-0.25em b}\kern-0.8em\TeX}}}

\begin{document}

\title{Clarifying the Compass: A Reflexive Narrative on Entry Barriers into HCI and Aging Research}

\author{Tianyi Li}
\email{li4251@purdue.edu}
\author{Jin Wei-Kocsis}
\email{kocsis0@purdue.edu}
\affiliation{%
  \institution{Purdue University}
  \country{USA}
}

\renewcommand{\shortauthors}{Li and Wei-Kocsis}

\begin{abstract}
This manuscript presents the perspectives and reflections of two researchers who were not previously engaged in aging research, regarding the gaps and barriers related to interdisciplinary collaboration on HCI and Aging research. The manuscript has two sections. In the first section, the authors discuss their observations on the disconnect between the needs of aging populations and the design of emerging technologies. The second section delves into their personal journey of developing empathy and a deeper understanding of older adults by volunteering in a senior living community, and shares their reflective thoughts on these experiences. 
\end{abstract}

\maketitle

\section{Introduction: The Need for Interdisciplinary Efforts Beyond HCI and Aging}
The nuanced needs and perspectives of older adults frequently go underrepresented in the design and development of human-AI systems~\cite{mannheim2019inclusion, 10.1145/3461702.3462609, pirhonen2020these}. As the global demographic shifts towards an aging population, it becomes imperative to ensure that technological innovations are not only accessible but also cater to the needs of individuals at all stages of life. Older adults form a diverse and growing segment of technology users, whose interaction with digital tools is shaped by a rich tapestry of social and cultural backgrounds~\cite{betts2019there, 10.1145/3479506, 10.1145/2556288.2557124}, health statuses~\cite{10.1145/3568162.3576995, 10.1145/2504335.2504344}, and personalities~\cite{zager2023assistive}. By prioritizing the inclusion of older adults in the digital landscape, we not only address their specific needs but also foster intergenerational connections~\cite{10.1145/3479506} and promote enhanced awareness and literacy in social justice and equity among people of all stages of life.

Designing human-computer interaction (HCI) to cater effectively to users across all life stages is a formidable task. On the technological front, the widespread integration of Artificial Intelligence (AI) is transforming the way human users engage with digital tools. 
On the human side, understanding the needs and pain points of different older adults populations requires cross-disciplinary training and long-term commitment. For example, technologies like smart home systems promise enhanced safety and autonomy, but may neglect the essential human values of privacy and dignity~\cite{chung2016ethical}.
The adoption of AI-enabled devices by older adults is further complicated by obstacles such as the perceived complexity of these technologies and a general apprehension towards new digital tools~\cite{goodarzi2024experiences}. A lack of intuitive design that aligns with their established routines and expectations can exacerbate these barriers. Moreover, the digital divide, deepened by factors like socioeconomic status and educational disparities, poses additional challenges, placing a subset of older adults at a distinct disadvantage in navigating the digital landscape~\cite{huang2024research}.

Below, the authors discuss their attempts to develop empathy and understanding of older adults from a local residential community, examining the challenges encountered and identifying the training requirements that may benefit other researchers from diverse disciplines aiming to contribute to HCI and Aging research.

\section{Reflexive Narratives}
In this section, the authors, initially unfamiliar with the older adult demographics, recount their path to gaining insight and empathy for the lives of older adults. Their objective was to explore the intersection of Human-Computer Interaction (HCI) and Aging research with their own areas of expertise, aiming to pinpoint more specific discrepancies between existing practices and the potential of emerging technologies. Although the primary intention was to identify needs and challenges addressable by their unique expertise, the authors encountered unforeseen personal limitations in framing research questions, applying values in practical research settings, and managing their own well-being. By sharing these reflective accounts, the authors underscore the hurdles faced by researchers venturing into the HCI and aging domain without prior experience, emphasizing the necessity for resources that enable such researchers to self-educate, self-regulate, and sustain their research endeavors.

\subsection{Researcher Backgrounds}
The two authors bring diverse disciplinary perspectives and hail from different stages in their careers, boasting varied technical and academic backgrounds. However, both lacked experience in conducting research with older adult populations prior to this study.

The first author, an early-career researcher, specializes in Sensemaking, Crowdsourcing, Visual Analytics, and Human-AI Interaction within the Human-Computer Interaction (HCI) field. Her previous research predominantly involved working adults adept at using technology in various capacities, leading to certain presumptions that might not extend to older adults. A significant focus of her work has been exploring how novice crowds can collaboratively interpret complex textual data and unravel mysteries to reach expert-level analysis. These novice participants are often sourced from platforms like MTurk, implying inherent traits such as proficiency in computer use and cognitive capabilities. In her other projects targeting different user demographics, the participants were typically domain experts involved in the design or development of technologies. Notably, none of the participants in her earlier studies were over the age of 65.

The second author, an mid-career researcher, has extensive research experiences in trustworthy machine learning, security and privacy, blockchain technologies, and cyber-physical-social systems, such as Internet of Things (IoT) systems, smart energy systems, wireless communication systems, workforce training systems, and other autonomous systems. A significant focus of her work has been developing and exploring intelligent cyber-physical-social systems that is reliable and trustworthy for the targeted applications even in the presence of uncertain and complex environment.

\subsection{Researcher Motivations}
The United States is on the verge of a demographic shift often described as a "silver tsunami," driven by the aging baby boomer generation and declining fertility rates. According to the Administration for Community Living, the U.S. population aged 65 and over is projected to reach 80.8 million by 2040 and 94.7 million by 2060. This substantial growth in the elderly population necessitates an urgent enhancement of senior healthcare systems. Without proactive interventions, the increasing number of older adults could lead to significant social and economic challenges, impacting everything from healthcare services to economic stability.  We would like to contribute to addressing the challenges from a technical perspective. 

Professionally, the relevance of aging research is magnified at the intersection of Human-Computer Interaction (HCI) and emerging human-AI systems. The global trend towards an aging population demands technological innovations that go beyond accessibility to truly accommodate the nuances of the aging process. By integrating AI into tools utilized by older adults, there is a significant opportunity to boost their independence, safety, and social connectivity, ensuring that technology serves as a bridge rather than a barrier in their daily lives.

However, on a personal level, the authors have observed the diverse challenges and limitations faced by older adults with varying health conditions and educational backgrounds as they interact with technology. Thus, our motivation is also fueled by firsthand experiences with aging family members. We hope of gain a deeper understanding of the disparities between academic research and the real-world impact of the AI solutions developed for teh aging populations.

\subsection{Researcher Presumptions}
Initially, the two authors have the following presumptions about aging population as users of AI-infused technologies:

Assumption 1: Emerging technologies have great potential to significantly enhance the efficiency of the healthcare system, making it more responsive and effective, particularly in handling the complex needs of an aging population.

Assumption 2: Given the high vulnerability of the aging population, it is essential to determine how to adapt emerging technologies to provide healthcare services effectively without compromising the quality of care.

Assumption 3: For older adults in good health, their interactions with technology are primarily influenced by their prior knowledge of technology, personality traits, and social circles. These factors play a significant role in shaping their perceptions and reactions to the design challenges they encounter while using technology. 

Assumption 4: Older adults in good health primarily derive benefits from AI technology through enhanced connections with family members and emotional support. However, there is a complex interplay between their personal preferences and the need to engage with the younger generation's world. 

\subsection{Initial Attempts to Understand Older Adults through Volunteer Work}
The authors engaged in volunteer work at a local senior living community. This community provides a range of services, from independent living options to more comprehensive nursing care, and is home to individuals with varying degrees of cognitive abilities, including those with dementia. 

The senior living community, henceforth referred to as "the community," welcomes local volunteers to assist with a variety of resident activities. Initial volunteering opportunities are available to all interested parties, allowing them to contribute to the community's monthly planned activities ranging from flower arranging to group exercises. After participating in two volunteering sessions, individuals wishing to continue their involvement must undergo a comprehensive background and health screening process. Volunteer responsibilities usually include assisting with residents' transitions between activities, collaborating with caregivers to facilitate events, and engaging in friendly conversations with residents.

\subsection{Reflections}
\textbf{Assumptions based on appearance during inter-personal interactions.} The HCI researcher's prior focus on technology interactions with working-age adults who were predominantly healthy and proficient in technology use subtly shaped my assumptions about older adults' capabilities and needs. A significant realization from volunteering at the senior living community was gaining a more nuanced understanding of residents' engagement levels. She occasionally misjudged individuals' energy and cognitive abilities based on initial impressions of their physical appearance and behavior, leading to unintentional biases in assessing their capacities. Observing and learning from the interactions between caregivers and the elderly was particularly eye-opening. Similarly, the second author, with less prior experience in human-centered research, found resonance in these observations.

\textbf{Emotional stress of expressing and receiving care.} In addition, both researchers' inexperience with this demographic made them overly cautious in tasks like maneuvering wheelchairs, potentially at the risk of affecting residents' self-esteem by treating them differently. This delicate balance between offering assistance and preserving their dignity was challenging. Group activities sometimes inadvertently treated those with lower cognitive health like children, yet private conversations revealed a contrasting depth of cognitive engagement and their desire not to burden others. This highlighted the internal conflict they face, needing support but feeling guilty about receiving it. 

\textbf{Foreseeable challenges in applying typical HCI research methods.} Adapting design methodologies for older adults with diminished cognitive abilities proved challenging~\cite{wang2019co}. Their shorter attention spans and difficulty in sustaining meaningful conversations raised questions about how to respect their autonomy while addressing their needs effectively. Engaging as a volunteer and fostering long-term relationships with the community and its members, while demanding significant time and ethical consideration, can be an effective way of gaining a profound understanding of their context and needs. This necessitates further scholarly discourse to inform and guide the development and conduct to more novel and effective ways to engage target audience with cognitive limitations. 

\textbf{Operationalizing research outcomes to practice.} The dedication of caregivers was evident in their endless patience and affection, crucial for maintaining a positive environment despite repetitive questions and the need for inclusive communication. The challenge for caregivers in managing multiple residents simultaneously, recognizing and accommodating their distinct needs, and responding to unforeseen situations, was palpable. Their personalized engagement with each resident, based on a deep understanding of individual personalities and conditions, was a testament to their commitment and highlighted the complexity of providing tailored support in such settings. Although there has been significant research and development in technologies aimed at facilitating caregiving tasks over the past decade~\cite{hossain2012human, zwierenberg2018lifestyle}, enabling various caregiving communities to adopt and tailor these technologies to their distinct contexts necessitates further and comprehensive efforts in HCI and interdisciplinary collaboration. 

\textbf{Barriers for adopting emerging technology in aging care.} 
In prior research endeavors, a range of emerging technologies including remote monitoring systems~\cite{9379817}, augmented and virtual realities~\cite{tsao2019development}, and robotic aides~\cite{alcover2021aging} have been developed, targeting diverse application areas with the potential to significantly empower caregivers in the aging sector. Despite their potential, our observations from volunteer work within senior living communities reveal a noticeable hesitancy towards the adoption of such technologies. This reluctance can primarily be attributed to challenges related to usability, alongside concerns regarding privacy and security that these technologies may pose.
Addressing these barriers necessitates a comprehensive investigation into the root causes that deter their adoption. It is imperative to engage in a collaborative effort involving caregivers, researchers, and stakeholders from various disciplines to explore these challenges in-depth. 

\bibliographystyle{ACM-Reference-Format}
\bibliography{ref}

\end{document}